\documentclass[11pt,a4paper]{article}

 \newcommand{\be}{\begin{equation}}
 \newcommand{\ee}{\end{equation}}
 \newcommand{\bl}{\begin{equation}\begin{array}{ll}}
 \newcommand{\el}{\end{array}\end{equation}}
 \newcommand{\bll}{\begin{equation}\begin{array}{lll}}
 \newcommand{\bdm}{\begin{displaymath}}
 \newcommand{\edm}{\end{displaymath}}
 \def\bea{\begin{eqnarray}}
 \def\eea{\end{eqnarray}}
 \def\barr{\begin{array}}
 \def\earr{\end{array}}
\def\p{\partial}

\def\half{\frac{1}{2}}

\def\third{\frac{1}{3}}
\def\2third{\frac{2}{3}}
\def\4third{\frac{4}{3}}
\def\3quart{\frac{3}{4}}

\def\sixth{\frac{1}{6}}

\def\lim{\rightarrow}

\def\bV{\bar{V}}

\def\cL{{\cal L}}

\def\dx{\dot{x}}

\def\ddx{\ddot{x}}

\def\drho{\dot{\rho}}

\def\dsig{\dot{\sigma}}
\def\dpsi{\dot{\psi}}
\def\dal{\dot{\alpha}}
\def\dbe{\dot{\beta}}
\def\dga{\dot{\gamma}}
\def\dpsi{\dot{\psi}}
\def\dA{\dot{A}}

\def\hba{\hat\textbf{a}}

\def\hga{\hat{\gamma}}

\def\ha{\hat{a}}

\def\Lag{{\cal L}}

\begin{document}
\raggedbottom

\title{{\bf Affine generalizations of gravity \\
 in the light of modern cosmology}}

\author{A.T.~Filippov \thanks{Alexandre.Filippov@jinr.ru}~ \\
{\small \it {$^+$ Joint Institute for Nuclear Research, Dubna, Moscow
Region RU-141980} }}

\maketitle

\begin{abstract}
 We discuss new models of an `affine' theory of gravity in
 multidimensional space-times  with \emph{symmetric connections}.
 We use and develop ideas of Weyl, Eddington, and Einstein,
 in particular, Einstein's proposal
 to specify the space - time geometry by use of the Hamilton
 principle. More specifically, the connection coefficients are
 determined using a `geometric' Lagrangian that is an arbitrary
 function of  the  generalized (non-symmetric)  Ricci curvature
 tensor (and, possibly, of other  fundamental  tensors) expressed in
 terms of the connection coefficients regarded as independent
 variables. Such a theory supplements the standard Einstein gravity
 with \emph{dark energy} (the cosmological constant, in the first
 approximation), a neutral massive (or tachyonic) vector field
 (\emph{vecton}), and massive (or tachyonic) \emph{scalar fields}.
 These fields couple only to
 gravity  and can generate dark matter and/or inflation. The new field
 masses (real or imaginary) have a geometric origin and must
 appear in any concrete model. The concrete  choice of the geometric
 Lagrangian determines further details of the theory, for example,
 the nature of the vector and scalar fields that can describe
 massive particles, tachyons, or even `phantoms'.
  In `natural' geometric theories,
 which are discussed here, dark energy must also arise.
     We mainly focus on intricate relations between geometry
     and dynamics while only very briefly considering
     approximate cosmological models inspired by the geometric
     approach.

\end{abstract}

\section{Introduction}
 Attempts to modify Einstein's general relativity began
 immediately after the general relativity was formulated in its
 final form (1915 -1916). Einstein himself
 added the cosmological constant term $\Lambda$ to save
 his static cosmology. After Friedmann's work (1922-1924) this
 modification was becoming more and more dubious.
 In 1918, Weyl
 developed a much more serious modification aimed at `unifying'
  gravity and electromagnetism
  (it is most clearly summarized in \cite{Weyl}).
  Starting from Levy-Civita's ideas on non-Riemannian
  connections (1917), he developed a theory of a manifold
  equipped with the connection that depends both on
  a metric tensor and on a vector field, which he attempted
  to identify with
 the electromagnetic potential.
 To get a consistent theory, Weyl introduced
 a general idea of gauge invariance which survived although the theory
 itself failed as he admitted later. At the same time, he invented
 conformal transformations and the conformal curvature tensor
 (Weyl's tensor).

 In 1919, Eddington proposed a more radical modification of general
 relativity \cite{Ed1}, \cite{Ed}. His idea was to start with
 a pure affine
 formulation of the gravitation, i.e. to first use the general
 symmetric affine  connection and only at some later stage
           to introduce a metric tensor.
 Eddington proposed several interesting ideas, including
 Eddington's density discussed below, but he could not construct
 a logically consistent theory based on an affine connection.

 Einstein, starting from Eddington's ideas developed a consistent
 theory based on using the Hamilton principle.
 His theory was presented in three beautiful and concise papers
 \cite{Einstein1}, later summarized by Eddington \cite{Ed1} and
 Einstein \cite{Einstein2} and soon forgotten (but see brief
 discussions  in \cite{Erwin}, \cite{Pauli}).
  The Einstein models were recently reinterpreted
 in \cite{ATF} and \cite{ATFn}, where
 a brief summary of the ideas and results in those papers that are
 of interest for modern investigations can be found.

 Einstein's key idea was to derive the concrete form of the
 affine connection by applying the Hamilton principle to
 a generic Lagrangian depending on the generalized Ricci curvature.
 This assumption completely fixes a geometry,
 which does not coincide with Weyl's geometry,
 but belongs to the same simple class
 recently introduced and discussed in \cite{ATF}.
 Einstein's unusual result was difficult to comprehend
 in twentieth of the last century and it remains somewhat
 puzzling these days.
 From the modern mathematics viewpoint, its origin could
 be ascribed to a sort of
        a mismatch between the  affine connection geometry
        and the Lagrangian `geometry'.
 At the moment, it is difficult to find a more detailed explanation.
 Possibly, this is an interesting mathematical problem.

 A more detailed presentation of the main ideas and results
      discussed here
 can be found in \cite{ATF} and \cite{ATFn},
      but we warn the reader that there are several
      essential changes in our approach to some problems.
 In Section~2 we give a very brief summary of the affine
 connection theory. In Section~3 we expose Einstein's
 approach, formulate  general geometric and physical
 requirements to the theory
 and their simplest realization.
 In Section~4 we discuss simple cosmological models.

\section{Geometry}
 Weyl's basic idea was that unifying gravity and
 electromagnetism requires using a non-Riemannian
 {\emph symmetric connection}. In general, the connection
 coefficients can be expressed in terms of the
 Riemannian connection $\Gamma^i_{jk}$ and of
 an arbitrary  third rank tensor $a^i_{jk}$ that
 is symmetric in the lower indices
 \be
 \label{app1}
 \gamma_{jk}^i  \,=\, \Gamma_{jk}^i[g] + a^i_{jk} \,.
 \ee
 Here $g_{ij}$ is an arbitrary symmetric tensor and
 $\Gamma^i_{jk}[g]$ is its Christoffel symbol
 \be
 \label{app11}
 \Gamma_{jk}^i [g] \,=\, \half g^{il} (g_{lj,k} +
 g_{lk,j} -  g_{jk,l}) \,,
  \ee
  where the commas denote differentiations and
  $g_{ij} g^{jk} = \delta^k_i\,$.
 More precisely, for any symmetric connection, there exists
 a symmetric tensor $g_{ij}$ and a tensor
 $a^i_{jk} = a^i_{kj}$ such that (\ref{app1}) is satisfied.

 The curvature tensor can be defined
 without using any metric:
 \be
 \label{1}
 r^i_{jkl} = -\gamma^i_{jk,l} + \gamma^i_{mk} \gamma^m_{jl}
 + \gamma^i_{jl,k} - \gamma^i_{ml} \gamma^m_{jk} \,.
 \ee
 Then, the Ricci-like (but {\emph non-symmetric}) curvature tensor can
 be defined by contracting the indices $i, l$ (or, equivalently,
 $i, k$):
 \be
 \label{2}
 r_{jk} = -\gamma^i_{jk,i} + \gamma^i_{mk} \gamma^m_{ji}
 + \gamma^i_{ji,k} - \gamma^i_{mi} \gamma^m_{jk}
 \ee
 (we again stress that $\gamma^i_{jk} = \gamma^i_{kj}$
 but $r_{jk} \neq r_{kj}$). Using only these tensors and the
 antisymmetric tensor density, we can construct a quite
 rich geometric structure.\footnote{The geometry of symmetric
 and nonsymmetric connections is carefully reviewed in
 \cite{Eisen} and in \cite{Erwin}. }

 The symmetric part of the Ricci curvature $r_{ij}\,$,
 \be
 s_{ij} \equiv \half (r_{ij} + r_{j\,i}) \,,
 \label{3a}
 \ee
 and its antisymmetric part,
 \be
 \label{4}
 a_{ij} \equiv \half (r_{ij} - r_{j\,i}) =
 \half (\gamma^m_{jm,i}  - \gamma^m_{im,j}) \,,
  \qquad  a_{ij,\,k} + a_{jk,i} + a_{ki,j} \equiv 0\,,
 \ee
 have essentially different roles in geometry and in physics.
 The antisymmetric tensor $a_{ij}$ strongly resembles the
 electromagnetic field tensor and is actually related to
 the massive (or tachyonic) vector field (vecton),
 which is proportional
 to the vector $a_i \equiv a^m_{im}~$
 (or to $\gamma_i \equiv \gamma^m_{im}$).
 According to (\ref{app1}),
 \bdm
 a_i \equiv \gamma^m_{mi} - \Gamma_{mi}^m \equiv
 \gamma_i - \p_i \ln \sqrt{|g|} \,,
 \edm
 where $g \equiv \textrm{det}(g_{ij})$. Therefore the vectors
 $a_i$ and  $\gamma_i \equiv \gamma^m_{im}$
 differ by a gauge transformation,
 which played and important role in Weyl's theory, but it is
 not so important for us, at the moment. It is important that
 $a_{ij}$ can be simply
 expressed in terms of the vector $a_i$ (or $\gamma_i$)
 \be
 \label{4a}
 a_{ij} \equiv -\half (a_{i,j} - a_{j,i})
 \equiv  -\half (\gamma_{i,j} - \gamma_{j,i}),
 \ee

 Eddington's scalar density
 \be
 \label{3}
 {\Lag} \equiv \sqrt{ -\det(r_{ij})} \,
 \equiv \, \sqrt{ -r} \,,
  \ee
 which resembles the fundamental scalar density of the Riemannian
 geometry, $\sqrt{-\textrm{det}(g_{ij})} \equiv \sqrt{-g}$,
 is also an important geometric and physical object. Einstein
 used it as the Lagrangian in his first paper on affine model.
 Here, we discuss more general Lagrangians that can be
 obtained using densities constructed of $s_{ij}$, $a_{ij }$,
 and $a_k$.

  Introducing the covariant derivative $\nabla^{\gamma}_i$
  (with respect to the connection $\gamma$) we can rewrite
  the symmetric part of the curvature as
 \be
 \label{4b}
 s_{ij} =  - \nabla^{\gamma}_m \gamma^m_{ij} +
 \half (\nabla^{\gamma}_i \gamma_j + \nabla^{\gamma}_j \gamma_i) -
 \gamma^m_{ni} \gamma^n_{mj} + \gamma^n_{ij} \gamma_n \,.
 \ee
 Using the `metric' covariant derivative
 $\nabla^g_i \equiv \nabla_i$ we can rewrite $s_{ij}$
 in the form
 \be
 \label{app3}
 s_{ij} = R_{ij}[g] - \nabla_m a^m_{ij} +
 \half (\nabla_i \, a_j + \nabla_j \, a_i) +
    a^m_{ni} a^n_{mj} - a^m_{ij} a_m \,,
  \ee
 where $R_{ij}[g]$ is the standard Ricci tensor of
 a Riemannian space with the metric $g_{ij}$.

 For a general symmetric connection one can introduce
 the concept of the geodesic curve, the tangent vector to
 which is parallel to itself at every point of the curve.
 Eisenhart \cite{Eisen} calls these curves `paths', but
  we may also call them geodesic curves (or geodesics)
  because they directly generalize
  the geodesics of the Riemannian geometry. The equations
  for geodesic curves of any symmetric connection $\gamma^i_{jk}$
  can be written in the form
 \be
 \label{4c}
 \ddx^{\,i} +\, \gamma^i_{jk} \, \dx^j \, \dx^k = 0 \,,
 \ee
 where the dot denotes differentiating with respect to
 the so called `affine' parameter $\tau$ of the curve
 $x^i(\tau)$. Using the affine parameter we can compare
 the distances between points on the same curve.

 For a particular path, the affine parameter
 is unique up to an affine transformation
 $\tau \mapsto \tau^{\prime} = a \tau + b$.
 Each connection define the unique set of paths, but
 all symmetric connections
 \be
 \label{4d}
 \hga^i_{jk} = {\gamma}^i_{jk} +
 \delta^i_j \, \ha_k + \delta^i_k \, \ha_j  \,,
 \ee
 with an arbitrary vector $\ha_k$, define the same
 paths. The Weyl (conformal) tensor $W^i_{jkl}$
 of connection (\ref{4d}) is independent of $\ha_k$ while
 the Ricci tensor and its symmetric and antisymmetric
 parts are $\ha_i$-dependent
 (see \cite{Eisen} for more details).

 An interesting class of connections is
 \be
 \label{4e}
 \hga^i_{jk} = {\Gamma}^i_{jk} [g] +
 \delta^i_j \, \ha_k + \delta^i_k \, \ha_j  \,,
 \ee
 where ${\Gamma}^i_{jk}[g]$ is a Riemannian connection
 (the Christoffel  symbol of a symmetric tensor $g_{ij}$).
 The paths of the connection $\hga^i_{jk}$ coincide with
 the geodesics of ${\Gamma}^i_{jk}[g]$, but the Ricci tensor
 of $\hat{\gamma}$ is symmetric if and only if $\ha_i = \p_i \, \ha$
 with some scalar $\ha$. We see that connection (\ref{4e}) is
 maximally close to the Riemannian connection
 ${\Gamma}^i_{jk}[g]$
 and may be called a `geodesically
  Riemannian' (or $g$-Riemannian) connection.
 Weyl and Einstein studied more general connections that belong
 to the following class introduced in \cite{ATF}, \cite{ATFn}:
 \be
 \label{a3}
 \gamma^i_{jk} = \Gamma_{jk}^i [g] +
  \alpha ( \delta^i_j \, \ha_k +  \delta^i_k \, \ha_j) -
   (\alpha - 2\beta) g_{jk}\, \ha^i  \,,
 \ee
  where  $\ha^i \equiv g^{im} \ha_m$.
 The Weyl connection corresponds to $\beta = 0$ and
 the $g$-Riemannian connection, to  $\alpha = 2\beta$.
  Einstein derived the connection for the
  space-time dimension $D=4$, his result is
  $\alpha = -\beta = \sixth$ (we generalize it to any
  dimension in the next section).

 Using (\ref{app3}) it is easy to calculate the physically
 important expression for the symmetric part of the Ricci curvature.
 The terms linear in $A$ are equal to
 \be
 \label{app4}
 (\alpha + \beta) (\nabla_i \ha_j + \nabla_j \ha_i) +
 (\alpha - 2\beta)\, g_{ij} \nabla_m \ha^m  \,,
 \ee
 and the quadratic terms are
 \be
 \label{app5}
 \ha_i \ha_j \,\bigl[(\alpha - 2\beta)^2 -3\alpha^2 \bigl]
  \,+\,
 2 \,g_{ij} \ha^2 (\alpha - 2\beta) (\alpha + \beta) \,.
 \ee
  It is easily seen that the sign of the first term in (\ref{app5})
 can be positive or negative, but the second term in (\ref{app5})
 and  the linear terms in (\ref{app3}) are nonzero in the general case.

 Before we leave pure mathematics and turn to more physical
 problems, we should mention one of
 the characteristic properties  of
 symmetric connections. For applications
 of geometry to gravity, it is very important that at every
 point of the affine-connected space-time manifold there must exist
 a geodesic coordinate system, such that the connection coefficients
 are zero at this point. Using the above formulas it is easy to
 prove that such a coordinate system exists if and only if the
 connection is symmetric. For symmetric connections, the Fermi
 theorem about the existence of geodesic coordinates along
 the curves also holds
 (for the precise definitions and proofs see \cite{Eisen}).

 \section{From Geometry to Dynamics}
 Einstein's approach to constructing the
 generalized theory of gravity consists of two stages.
   In the first  stage, he assumed
   that the general symmetric connection should be
  restricted by the Hamilton principle for a general Lagrangian
 density depending either on $r_{ij}$ (in the first two papers)
 or on $s_{ij}$ and $a_{ij}$ separately
 (in the third paper).\footnote{
 This idea was quite alien to Weyl and Eddington, who
 began by formulating a particular geometry.
 They therefore postulated the connection (\ref{a3}) with
 $\beta = 0$   and then tried to
 write some equations generalizing the Einstein equations.
 Although the general approaches and
 the connection coefficients differed,
 the equations considered by the three authors have many
 features in common. In particular, the nonzero cosmological
 constant (exactly or in an approximation) and massive (tachyonic)
 vector field.}
 He gave no motivation for this assumption, but it is easy to see
 that the resulting theory in the limit $a_{ij}=0$  is
 consistent with the standard general relativity supplemented
 with a cosmological term. In this stage, Einstein succeeded in
 deriving the remarkable expression (\ref{a3}) for the connection
 (with $\alpha = -\beta = \sixth $)
 and the general expression for $s_{ij}$ depending on
 a massive (tachyonic) vector field and the metric
 tensor density $\textbf{g}^{ij}$.

 In the next stage, a concrete Lagrangian density
 ${\Lag}(s_{ij} \, , a_{ij})$  should be chosen.
 Einstein did not formulate any principle for selecting
 a Lagrangian, and both from geometric and physical standpoint
 his concrete choice seems sufficiently arbitrary,
 especially in the third paper. We believe that his best choice was
 made in the first two papers and, indeed, very similar
 effective Lagrangians are considered in modern applications
 of the superstring theory to cosmology.

 Let us try to formulate
 requirements for a geometric Lagrangian density ${\Lag}$:

 \medskip
 \noindent {\bf 1.}~It is independent of dimensional constants.

 \noindent {\bf 2.}~Its integral over the $D$-dimensional
 space-time is dimensionless.

 \noindent {\bf 3.}~It can depend on tensor variables having a direct
 geometric  meaning\\
 and a natural physical interpretation.

 \noindent {\bf 4.}~Most importantly, the  resulting
 generalized theory must agree with \\
 the well-established experimental consequences
 of the standard Einstein theory.

 \medskip
 \noindent The last requirement is rather difficult to check without
 a detailed development of the theory.
 It is not easy to really motivate the requirements
 \textbf{1-2}, and in fact they simply summarize some ideas
 of \emph{`naturalness'} that are naturally subjective.
  The first and the second requirement
 emphasize that in geometry we measure only lengths
 (we always use the condition that the speed of light $c=1$).
   Requirement \textbf{3}
 is somewhat vague and depends on our
 understanding of what is `geometry' and what is `physics'.
 Roughly speaking, before we use any Lagrangian we are in
 the domain of pure geometry. The `geometric Lagrangian'
 is a centaur that depends only on geometric variables, but
 defines a Lagrangian structure through the variational
 principle and allows to define physical variables.

 Clearly, the variables  $r_{ij}$,  $s_{ij}$, $a_{ij}$,
 and $a_k \equiv a^i_{ik}$
 satisfy requirement {\bf 3}.
 Likewise, we can take as a Lagrangian density
 Eddington's density  (\ref{3}), which is the simplest possible
 choice satisfying requirements \textbf{1-3}.
 It is easy to see that there exists a wide class of
 scalar densities also satisfying  \textbf{1-3}.
 In the first two papers of \cite{Einstein1},
 Einstein discussed the geometric Lagrangian defined
 by Eddington's density (\ref{3}).
 In the third paper, he assumed that one can take any Lagrangian
 depending on $s_{ij}$ and $a_{ij}$ as independent variables,
 i.e.
 \be
 {\Lag} \, = \, {\Lag} (s_{ij} , a_{ij}) \,,
 \label{5}
 \ee
 but considered a concrete Lagrangian not satisfying
 requirements  \textbf{1-2}, which in fact is very
 similar to Weyl's Lagrangian studied in \cite{Weyl}.
 A simplest generalization of Eddington-Einstein
 Lagrangian that depends on  $s_{ij}$ and $a_{ij}$
 and satisfies all the requirements \textbf{1-3}
 was proposed in \cite{ATFn}
 \be
 \label{3aa}
 {\Lag} = {\Lag} (s_{ij} + \nu a_{ij}) =
 \sqrt{-\det(s_{ij} + \nu a_{ij})},
 \ee
 where we take the minus sign because $\det(s_{ij})<0$
 (due to the local Lorentz invariance requirement) and we
 naturally assume that the same is true for
 $\det(s_{ij} + \nu a_{ij})$ (to reproduce Einstein's
 general relativity in the limit
 $\nu a_{ij} \rightarrow 0$).\footnote{
 In fact, we introduce this new parameter to
 disentangle the scale of the mass parameter
 of the vector field from the cosmological
 constant. It turns out that for $\nu = 1$, i.e.,
 for original Eddington - Einstein Lagrangian
 (\ref{3}), the mass parameter is
 close to $\sqrt{\Lambda}$ (see \cite{ATFn}).
 }

 This Lagrangian can be further generalized if we consider
 other scalar densities of the weight two constructed of
 $s_{ij}$, $a_{ij}$, and $a_i$. The basic element of the
 construction is the density
  \be
  \label{3b}
 d_0 \equiv 4! \, {\det}(s_{ij}) = \epsilon^{ijkl}
   s_{im} s_{jn} s_{kr} s_{ls} \epsilon^{mnrs} \, \equiv \,
 \epsilon \cdot s \cdot s \cdot s \cdot s \cdot \epsilon  \,.
 \ee
 Using the same natural notation we introduce the densities
 depending also on $a_i$ and $a_{ij}\,$:
 \be
  \label{3c}
  d_1 \equiv
 \epsilon \cdot s \cdot s \cdot s \cdot {\bar a} \cdot
 \epsilon \,,  \qquad
 d_2 \equiv  \epsilon \cdot s \cdot s \cdot a \cdot a \cdot
 \epsilon \,, \qquad
 d_4 \equiv \epsilon \cdot a \cdot a \cdot a \cdot a \cdot
 \epsilon \,.
 \ee
 where $d_4 = 4! \, \det (a_{ij})$ and
 ${\bar a}$ denotes the matrix $a_i a_j\,$.
 It is easy to find that
 \be
  \label{3d}
 {\det(s_{ij} + \nu a_{ij})} =
 \frac{1}{4!} \, (d_0 + 6\nu^2 d_2 + \nu^4 d_4) \,.
 \ee

 Then a more general geometric Lagrangian density can be written,
 \be
 \label{new1}
 {\Lag} \equiv \alpha_0 \, \sqrt{| \, d_0  \,+
 \, \alpha_1 d_1 \,+\, \alpha_2 d_2 \,+\,
 \alpha_4 d_4 |} \,\,,
  \ee
 where $\alpha_i \in \Re$ and $d_0 < 0$.
 This Lagrangian is in principle as good as
 the simplest Lagrangians (\ref{3}) or (\ref{3aa}),
 although working with it is more difficult
 and the spectrum of physical models described by it
 is much wider.
 Depending on the signs of the numerical coefficients, we
 could then obtain a positive or a negative cosmological constant
 and also the standard or the exotic (phantom) sign of the vector
 field kinetic energy. We note that the Lagrangian
 has zeroes in the general case,
 like the simpler Born-Infeld Lagrangian.\footnote{
 As a matter of fact, the so called Born-Infeld Lagrangians
 (see \cite{Born} -  \cite{Langlois}, etc.)
  are famous descendants of the forgotten
  Eddington-Einstein Lagrangian.
 }

 We note that the Lagrangians (\ref{3}) and (\ref{3aa})
 are are written in the form independent of $D$,  although
 the analytic expression for the dependence of the
 determinants on  $s_{ij}$ and $a_{ij}$ essentially
 depends on $D$. In \cite{ATFn}, we proposed
 as one of the requirement a formal independence of the
 geometric Lagrangian on $D$. Here we omit this requirement
 because it was poorly formulated and essentially restricts
 possible Lagrangians to (\ref{3}) and (\ref{3aa}) or, at most,
 allows to extend $\det(s_{ij} + \nu a_{ij})$ to
 $\det(s_{ij} + \nu a_{ij} + \bar{\nu} a_i a_j)$.

 The starting point for Einstein (in his first paper
   of the series  \cite{Einstein1})
 was to write the action principle and to assume that (\ref{3})
 is the Lagrangian density depending on 40 connection functions
 $\gamma^i_{jk}$.
 Varying the action with respect to these functions,
 he derived 40 equations that allowed him to find the
 expression for $\gamma^i_{jk}$ given by (\ref{a3}) with
 $\alpha = - \beta = \sixth \,$.

 We reproduced the main steps of the proof in \cite{ATFn}.
 Here, we somewhat generalize the derivation
 to an arbitrary dimension $D$ and assume that
 the geometric Lagrangian depends also on
 $\gamma_i \equiv \gamma^m_{im}$.
 We define the new tensor densities\footnote{
 Following Eddington's notation, we let boldface Latin letters
 denote tensor densities.
 The derivatives in (\ref{10}) and (\ref{10a})
 must be properly symmetrized,
 which is easy in concrete calculations.
  We tacitly assume that geometry
  has only a single dimensional constant,
 e.g., the cosmological constant $\Lambda$ with the dimension
 $L^{-2}$. Possibly, the characteristic constants for the
 symmetric and antisymmetric parts differ.
 To restore the correct dimension in (\ref{10}) and (\ref{10a}),
 we must then multiply the densities $\textbf{g}^{ij}$ and $\textbf{f}^{ij}$
 by $\Lambda_s$ and $\Lambda_a$. The simplest example of this
 asymmetry is suggested by Lagrangian (\ref{3aa}).
 }
  \be
 \label{10}
  {{\p {\Lag}} \over {\p s_{ij}}} \equiv \textbf{g}^{ij} \,,
  \qquad
 {{\p {\Lag}} \over {\p a_{ij}}} \equiv \textbf{f}^{ij} \,,
  \qquad
 {{\p {\Lag}} \over {\p \gamma_i}} \equiv \textbf{b}^i \,,
  \ee
  and introduce a conjugate Lagrangian density
  ${\Lag}^* \, = \, {\Lag}^* (\textbf{g}^{ij} ,
  \textbf{f}^{ij} ,\textbf{b}^i)$
  by a Legendre  transformation,
 \be
 \label{10a}
 s_{ij} = {{\p {\Lag}^*} \over {\p \textbf{g}^{ij}}} \,, \qquad
 a_{ij} = {{\p {\Lag}^*} \over {\p \textbf{f}^{ij}}} \,, \qquad
 \gamma_i = {{\p {\Lag}^*} \over {\p \textbf{b}^i}} \,.
   \ee
   By varying ${\Lag}$ in $\gamma^i_{jk}$ and using
   the above definitions, we can then show that the condition
   $\delta {\Lag} / \delta \gamma^i_{jk} = 0$ yields the
   following 40 equations
   \be
   \label{10b}
  2 \nabla^{\gamma}_i \, \textbf{g}^{jk} \,= \,\,
  \delta^k_i \, [\nabla^{\gamma}_m \, (\textbf{g}^{jm} +
  \textbf{f}^{jm}) - \textbf{b}^j ] +
 \delta^j_i \, [\nabla^{\gamma}_m \, (\textbf{g}^{km} +
 \textbf{f}^{km}) - \textbf{b}^k ] \,,
   \ee
 where $\nabla^{\gamma}_i$ is the covariant derivative with respect to the
 affine connection $\gamma$. Using the expression for the covariant
 derivative of the tensor density $\textbf{f}^{jk}$,
 \be
 \label{10c}
 \nabla^{\gamma}_i \, \textbf{f}^{jk} \,= \,\,
 \p_i \, \textbf{f}^{jk} + \,\gamma^j_{im} \, \textbf{f}^{mk} +\,
 \gamma^k_{im} \, \textbf{f}^{jm}   -
 \gamma^m_{im} \, \textbf{f}^{jk} \, ,
 \ee
  we find that  $\nabla^{\gamma}_i \, \textbf{f}^{ki} =
  \p_i \, \textbf{f}^{ki}$.
  Then, defining the vector density $\hat\textbf{a}^k$ by
 \be
 \label{10d}
 \p_i \, \textbf{f}^{ki} \,-\, \textbf{b}^k
 \, \equiv \, \hba^k ,
  \ee
  we easily find that
 \be
   \label{10e}
  \nabla^{\gamma}_i \, \textbf{g}^{ik} \,= \,\,
  - {\frac{D+1}{D-1}} \hba^k  \,,
   \ee
 and hence
 \be
   \label{10f}
  \nabla^{\gamma}_i \, \textbf{g}^{jk} \,= \,\,
  - {\frac{1}{D-1}} \bigl(\delta_i^j \hba^k +
  \delta_i^k \hba^j \bigr).
   \ee
 Defining the Riemann metric tensor
 $g_{ij}$ by the equations
 \be
 \label{11}
 g^{ij} \sqrt{-g} \, = \, \textbf{g}^{ij} \,, \quad
 g_{ij} \, g^{jk} \, = \, \delta^k_i \,,
\ee
 we can then define the corresponding Riemannian
 covariant derivative $\nabla_i$, for which\,
 \be
 \label{11a}
 \nabla_i \, g_{jk} = 0, \quad
 \nabla_i \, g^{jk} = 0 .
  \ee

 Taking the above into account, we can now use (\ref{10f})
 to derive the expression  for $\gamma_{jk}^i$
 in terms of the metric tensor $g_{ij}$ and
 of  the vector $\ha^k \equiv \hba^k/\sqrt{-g}$,
 \be
 \gamma_{jk}^i  \,=\, \Gamma_{jk}^i[g] \,+\,
 \alpha_D \, \bigl[\, \delta^i_j \, \ha_k +
 \delta_k^i \, \ha_j - (D-1)\, g_{jk} \, \ha^i \bigr] \,,
 \label{a31}
 \ee
 which corresponds to $\alpha = \alpha_D$  and
  $\beta = \beta_D$ in (\ref{a3}), with
 \be
 \label{a32}
 \alpha_D \equiv [(D-1)(D-2)]^{-1} \,,
 \qquad
 \beta_D \equiv - [2(D-1)]^{-1}\,.
 \ee
 This coincides with Einstein's result for $D=4$
 and never gives Weyl's or $g$-Riemannian connection.

  We cannot go deeper into discussions of further relations
 between geometry of affine connections and dynamical
 principles. But the above results show that these
 relations are rather complex and we do not yet
 understand their nature. Having in mind that for different
 connections the physical models may be drastically
 different, we tried to add new natural variables
 into the geometric Lagrangian. However, the
 class of connections obtained as an output of
 Einstein's approach did not change at all!
 It can be argued that there are many other, not yet
 explored options, but at the moment, we do not even know
 how to obtain Weyl's or $g$-Riemannian connections
 following Einstein's approach.

 One of the possibilities is to abandon some of
 Einstein's assumptions. The most serious drawback
 (or virtue, depending on a viewpoint) of his approach
 is that two  pairs of the basic variables of the theory,
 ($s_{ij}$,  $\textbf{g}^{ij}$) and
 ($a_{ij}$,  $\textbf{f}^{ij}$),
 having very different geometrical and physical
 meaning, are treated symmetrically. Definition
 (\ref{10}) looks quite natural for the metric
 density because Einstein's Lagrangian for the pure
 gravity theory is simply $\textbf{g}^{ij} R_{ij}$.
 But, Einstein's definition of $\textbf{f}^{ij}$, in fact,
 tacitly assumes that the geometric Lagrangian is
 independent of $\gamma_i$ or  $a_i$ (accordingly,
 he did not introduce the vector density
 $\textbf{b}^i$).\footnote{
 This may look rather paradoxical, but, as we have seen,
 the mass term is dictated by the geometry, and its
 trace, the term $\sim a_i a_j$, is already present in the
 expression for $s_{ij}$. The physical mass term itself is coming
 into being when we write an effective physical Lagrangian.}
  Then it follows that the connection
 structure is given by (\ref{a3}), where $g_{ij}$ is
 arbitrary and $\ha_k$ is proportional to $a^i_{ik}$.
 Moreover,
 $\textbf{f}_{ij} \equiv g_{im} g_{jn} \textbf{f}^{mn}$
 is proportional to $a_{ij}$. However,
 in the Einstein approach,
 the vector $a_i$ and the tensor $a_{ij}$ related by
 Eq.(\ref{4a}) (which exists in the most general geometry)
  are reproduced with the aid of rather
 indirect and complex relations $(\ref{10})$
 and $(\ref{10d})$.

  One more question about Einstein's approach concerns the role of
  the  metric tensor  $g_{ij}$ in geometry and in physics.
  In a more general sense,  this is a question
  about the meaning of Einstein's  geometric Lagrangian.
 In  Weyl's  geometric approach,  the tensor $g_{ij}$ is
 introduced  from the very
 beginning, but it is defined up to the Weyl transformations.
 In Einstein's approach, it is defined (uniquely,
 at a first glance)
 using the  Hamilton principle.  But we  know that
 the metric depends
 on the choice of the Weyl gauge (frame)  and  that
 the vector field also   depends on this choice.
  In particular,  the role of conformal transformations,
  the choice
 of the Weyl gauge (frame), and especially the significance
 and consequences of choosing different  independent fields
 in the initial geometric Lagrangian  must be carefully
 investigated  and understood.

\section{Models}
 There are many other questions, which should be
 carefully discussed, but we postpone the discussion
 to future publications. Here, we present a simple example
 demonstrating how to eventually pass from geometry
 to physics. Pure geometry gives us equations (\ref{4})
  and (\ref{app3}). With $a^i_{jk}$ given by
 (\ref{a3}), their right-hand sides are given by
 $(a_{i,j} - a_{j,i})/2$, where
 $a_i = (D\alpha +2\beta) \,\ha_i$, and by the sum
 of (\ref{app4}) and (\ref{app5}).\footnote{
 These expression are significantly simplified
 when $\alpha  + \beta = 0$ (Einstein's connection) or
 $\alpha  - 2\beta = 0$ (g-Riemannian connection.}
 To derive $s_{ij}$ and $a_{ij}$ in terms of the
 `physical' variables $g_{ij}$ and $f_{ij}$ we
 must choose a Lagrangian (e.g., (\ref{3aa})) and
 then solve equations (\ref{10}) with respect to the
 geometric variables $s_{ij}$ and $a_{ij}$.
 Alternatively, if we know the conjugate Lagrangian
 ${\Lag}^* (\textbf{g}^{ij} ,\textbf{f}^{ij} , ...)$,
 we can directly calculate them using (\ref{10a}).

 In \cite{ATFn}, we reproduced Einstein's result
 (see \cite{Einstein1} and \cite{Ed1}):
 \be
 \label{11e}
 {\Lag} \equiv \sqrt{ -\det(r_{ij})} \,=\,
 4\sqrt{-\det(\textbf{g}^{ij} + \textbf{f}^{ij})}
 \, \equiv \, 4\sqrt{-\det(g_{ij} + f_{ij})} \,
 =\, {\Lag}^*\,.
  \ee
 Actually, ${\Lag}^* = {\Lag}$ follows from the fact
 that ${\Lag}$ is a homogeneous function of the degree
 two, but the concrete expression for ${\Lag}^*$
 must be obtained by a direct calculation.
 Now we can show that the relation like (\ref{11e})
 holds also for Lagrangian (\ref{3aa}).
 For simplicity, we do this by direct
 calculations in a `dimensionally reduced' case.

 We first define a `spherical reduction' not using
 any metric. Suppose that $s_{ij}$ and $a_{ij}$
 are functions of $(x^0,x^1)$ and that $a_2 = a_3 =0$
 (therefore, only $a_{01} = -a_{10} \neq 0$).
 We then assume that the symmetric matrix has
 the following nonzero elements:
 $s_{ij} = \delta_{ij}\, s_i \,$, except
 $\,s_{01} = s_{10} \neq 0$ (our result will
 not change if also $s_{23} \neq 0$).
 Explicitly deriving $s_{ij} + \nu a_{ij}$, we can
 find $\textbf{g}^{ij}$ and $\textbf{f}^{ij}$
    (using (\ref{10})) and hence derive
 $\det(\textbf{g}^{ij} + \lambda \textbf{f}^{ij})$
  in terms of $s_{ij}$ and $a_{ij}$
 \be
 \label{11f}
 16 \det(\textbf{g} + \lambda \textbf{f}) =
 \det[s + (\nu^2 \lambda) \,a] .
 \ee
 It follows that for $\lambda \nu = 1$ we
 have
 \be
 \label{11g}
 {\Lag} = -\half \sqrt{|\det(s + \nu \,a)|} =
  -2\Lambda \sqrt{|\det(\textbf{g} +
  \lambda \textbf{f})|}
  = {\Lag}^* \,,
 \ee
 where we introduced the `cosmological' parameter
 $\Lambda$ having the dimension $L^{-2}$ (the sign
 and normalization are arbitrary chosen in relation
 to the cosmological interpretation).
 This result is written in the form not implying the
 spherical reduction, and we suppose it is true in the general
 four-dimensional theory. In arbitrary dimension ($D \neq 2$)
 it must be somewhat modified (see \cite{ATFn}).

 We considered the spherical reduction of
 the four-dimensional manifold. To demonstrate what
 can be obtained in higher dimension we consider
 a `spherically symmetric' five-dimensional model
 that is dimensionally reduced to dimension four.
 Let us add to the $4 \times 4$ matrix
 $r_{ij} = s_{ij} + \nu a_{ij}$ (satisfying all
 the above requirements) the anti-symmetric elements
  $a_{i4} = -a_{4i}\,$ ($i=0..3$), which also depend
  only on $(x^0,x^1)$. In addition, we assume that
 $a_2 = a_3 =0$ and that
 $s_{i4} = s_{4i} =  s_4 \, \delta_{i4}\,$,
 for $i=0,..,4$. Then we can immediately calculate
  $\det(s_{ij} + \nu a_{ij})$:
 \be
 \label{11h}
  \prod_{i=0}^4 s_i  \bigl[ 1 -
 (s_{01}^2 -\nu^2 a_{01}^2) (s_0 s_1)^{-1} +
 \nu^2 a_{04}^2 (s_0 s_4)^{-1} +
 a_{14}^2 (s_1 s_4)^{-1} -
 2\nu^2 s_{01} \, a_{04}\, a_{14} \, (s_0 s_1 s_4)^{-1}
 \bigr]   .
   \ee
 The terms containing $a_{04}^2$ and $a_{14}^2$
 should be interpreted as kinetic terms of
 the scalar field $a_4$ in four dimensions, because
 $a_{i4} = \p_i a_4/ 2$ while
 $a_{10} = (\p_1 a_0 - \p_0 a_1) / 2$ is
 the field tensor of the four-dimensional vector field.
 It can be seen that this scalar field is massive or tachyonic
 (in the simplest reduction, its mass coincides with
 that of the vecton).

 Taking the square root of the determinant
 (\ref{11h}) as a geometric Lagrangian, we can construct
 a two-dimensional model effectively describing spherically
 symmetric solutions of the four-dimensional gravity
 coupled to the vecton and scalar fields.
 By further reductions to static or cosmological solutions
 we can construct corresponding one-dimensional dynamical
 systems describing static states with horizons and
 cosmological models.
 The cosmological models look realistic enough because
 they incorporate a natural sources of the dark energy,
 inflation, and, possibly, some candidates for the dark
 matter (for a more detailed discussion see
 \cite{ATF}, \cite{ATFn})).
 We do not consider a general  theory and simply use
 for ${\Lag}^*$  expression (\ref{11g}).
 In the dimension $D$ we must slightly generalize
 (\ref{11e}) to
 \be
 {\Lag} \, \equiv \, \sqrt{ -\det(s_{ij} + a_{ij})}
 \, = \, \sqrt{-g} \ [-2^D \det (\delta_i^j +
 \lambda f_i^j)]^{1/(D-2)} = {\Lag}^* \,,
 \label{dr}
 \ee
 Following the relevant calculations of \cite{ATFn}
 we can write a `physical' Lagrangian
  \be
 \label{8}
 \Lag_{eff} =  \sqrt{-g} \,
 \biggl[ -2 \Lambda \,[\det(\delta_i^j +
 \lambda f_i^j)]^{1/(D-2)} +  R(g) +
 c_a \, g^{ij} a_i a_j \biggr] \,,
 \ee
  which should be varied with respect to the metric and the
  vector field; $c_a$ is a parameter depending on $D$
  (Einstein's first model is obtained for $D=4$ and
  $c_a = 1/6$).   When the vecton field is zero,
  we have the standard Einstein gravity
  with the cosmological constant.

  The theory described by (\ref{8}) is very complex,
  even at the classical level. Its spherically symmetric
  sector is essentially simpler, but the corresponding
  two-dimensional field theory is certainly not integrable.
  We do not know even how to construct
  its physically interesting approximate
  solution. Further dimensional reductions
  to one-dimensional static or cosmological theories
  also give non-integrable dynamical systems, though
  some approximate solutions can possibly be derived.
  A more effective is the small-field approximation
  formally equivalent to expanding (\ref{8}) in powers
  of $\lambda^2$. Keeping only the first-order
  correction we then obtain a nice-looking
  field theory:
 \be
 {\Lag}_{eff}\, \cong \,  \sqrt{-g}\,
 \biggr[R[g] - 2\Lambda -
 \kappa \biggr(\half F_{ij} F^{ij} + \mu^2 A_i A^i
 + g^{ij} \p_i \psi \, \p_j \psi + m^2 \psi^2 \biggl) \biggl]\,,
 \label{13a}
 \ee
 where $A_i \sim a_i$, $F_{ij} \sim f_{ij}$,
 $\kappa \equiv G/c^4$
 and we use
 the CGS dimensions.

 This simplified theory still keeps
 traces of its geometric origin: the simplest form of
 the dark energy (the cosmological constant $\Lambda$),
 massive (or tachyonic) vector and scalar fields, which
 can describe inflation and/or imitate
 dark matter.
 The most popular inflationary models require a few
 massive scalar particles usually called inflatons
 (see, e.g.,  \cite{Starobin} - \cite{Rubakov}).
 Without massive scalar fields there is no
 simple inflation mechanism with one massive
 vecton. However, with the tachyonic vecton (see \cite{Ford})
 or  with several massive vector particles,
 it is probably easier to find a realistic inflation models
 (see  \cite{Bertolami} - \cite{Germani}; some of
 these papers also discuss possible role of massive vector
 particles in dark energy and dark matter mechanisms).

   The simplest cosmology
 can be obtained by the naive reductions
 using the metric
 \be
 ds_4^2 = e^{2\alpha} dr^2 + e^{2\beta} d\Omega^2 (\theta , \phi) -
 e^{2\gamma} dt^2  \,,
 \label{eq1}
 \ee
 where $\alpha, \beta, \gamma$ depend on $t$
  and where  $d\Omega^2 (\theta , \phi)$
  is the metric on the 2-dimensional sphere
 $S^{(2)}$. Then  the cosmological
 reduction of four-dimensional
 theory (\ref{13a}) can be easily found.
 As was shown in \cite{ATFn}, the Lagrangian
 can be written in the form ($A \equiv A_z (t)$):
 \be
 \cL_c =
 e^{2\beta} \bigl[e^{-\alpha - \gamma} \dA^2 -
 e^{-\alpha + \gamma} \mu^2 A^2 -
 e^{\alpha +\gamma} (V + 2\Lambda)  -
 e^{\alpha - \gamma} (2\dbe^2 + 4\dbe \dal - \dpsi^2) \bigr] \,.
 \label{eq10}
 \ee

  To write the corresponding equations of motion in a
 most clear and compact form, we introduce the notation
 \be
 \rho \equiv \third (\alpha + 2\beta) \,, \quad
 \sigma \equiv \third (\beta - \alpha) \,, \quad
 A_{\pm} = e^{-2\rho + 4\sigma} (\dA^2 \pm
 \mu^2 e^{2\gamma} A^2) \,, \quad
 \bV \equiv V(\psi) + 2\Lambda \,.
 \label{eq11}
 \ee
 Then the exact
 Lagrangian for vecton-scalar cosmology is:
 \be
 \cL_c = e^{2\rho - \gamma} (\dpsi^2 - 6\drho^2 + 6\dsig^2) +
 e^{3\rho - \gamma} A_{-}  -
 e^{3\rho + \gamma} \, \bV(\psi) \,.
 \label{eq12}
 \ee
 We see that $A, \psi, \rho, \sigma$ are dynamical variables
 and $e^\gamma$ is a Lagrangian multiplier, whose variations
 yield  the energy constraint:
 \be
 \dpsi^2 - 6\drho^2 + 6\dsig^2 + A_{-} + e^{2\gamma} \,\bV = 0\,.
 \label{eq13}
 \ee
 As in any gauge theory with one constraint of this type,
 we can choose one gauge fixing condition. The standard
 conditions are $\gamma = 0$ or $\gamma = \alpha$.

  The other equations are
 \be
 \ddot{A} + (\drho + 4\dsig - \dga) \dA +
 e^{2 \gamma} \mu^2 A = 0 \,,
 \label{eq14}
 \ee
  \be
 4\ddot{\rho} + 6\drho^2 -4\drho \dga  - 6\dsig^2 + \third A_{-}
 + \dpsi^2 - e^{2 \gamma} \, \bV = o\,,
 \label{eq15}
 \ee
 \be
 \ddot{\sigma} + 3 \dsig \drho - \dsig \dga -\third A_{-} = 0 \,,
 \label{eq16}
 \ee
 \be
 \ddot{\psi} + (3\drho - \dga)\dpsi +
 \half e^{2 \gamma} \, \bV_{\psi}  =  0 \,,
 \label{eq17}
 \ee
 The system can be simplified by applying a gauge fixing
 condition. In particular, it can be written in
  a more standard form
 by excluding `dissipative' terms from equations for
 $\rho$, $\sigma$, and $\psi$. This can be done by choosing
 the gauge $\gamma = 3\rho$. In this gauge, the equations of
 the vecton cosmology are easier to compare with integrable
 dynamical systems that were extensively studied in cosmology
 and in black hole theory
 (see, e.g. \cite{CAF1} - \cite{ATF7}).

 Unfortunately, in any gauge these equations remain
 much more complex than the equations
 of the scalar cosmology. They are not integrable
 in any sense and rather difficult for a qualitative analysis.
 They would be greatly simplified if we could  neglect the
 $\sigma$ field. Unfortunately, this is obviously impossible
 in general,  because then $A_{-}$ would be zero and the last
 condition would be incompatible with the other equations.
 This means that the exact solutions of the model (even with
 many scalar fields minimally coupled to gravity)
 must be non-isotropic.\footnote{
 If we introduce other scalar fields nonminimally coupled to
 gravity, then this statement may become not valid. At the
 moment, we are not ready to add other vector fields or
 fields with the spin $1/2$.}

 In conclusion, we note that the geometrical and
 dynamical models discussed in this paper are not well understood,
 both conceptually and technically. Much work on them should
 be done before a realistic cosmological model could be
 constructed.


 {\bf Acknowledgment:}
 It is a great pleasure to dedicate this article to
 a very old friend, Andrei Slavnov. I am happy to know him
 for many years and to share his deep vision of science and
 life. I hope to have many opportunities to see him and to
 learn from him in future.

  This work was supported in part by the Russian Foundation
 for Basic Research (Grant No. 09-02-12417 ofi-M)


 \end{document}